\documentclass[11pt]{article}
\usepackage{graphicx}
\usepackage{amsmath}
\usepackage{amsfonts}
\usepackage{amssymb}
\usepackage{wrapfig}
\usepackage{hhline}

\def\be{\begin{eqnarray}}
\def\ee{\end{eqnarray}}
\def\nn{\nonumber}

\def\Hm{\mathcal{H}}
\def\Km{\mathcal{K}}
\def\Rm{\mathcal{R}}
\def\Sm{\mathcal{S}}

\def\Hfr{\mathfrak{H}}
\def\Kfr{\mathfrak{K}}

\def\gC{\grave{C}}
\def\ggC{\grave{\grave{C}}}

\def\Kh{\mathtt{Kh}}

\def\Tor{\mathtt{Tor}}
\def\Tw{\mathtt{Tw}}

\def\Sh{\mathtt{Sh}}

\def\sym{\mathtt{sym}}
\def\asm{\mathtt{asm}}
\def\adj{\mathtt{adj}}
\def\trv{\mathtt{trv}}

\def\sgn{\mathtt{sgn}}
\def\Cr{\mathtt{cr}}

\def\dms{\diamondsuit}
\def\dm{\diamond}

\begin{document}

\title{{\Large {\bf
      Khovanov polynomials for satellites and asymptotic adjoint polynomials}
  	\vspace{-1cm}
    \date{}
    \author{
      {\bf A. Anokhina$^{a,*}$}
      {\bf A. Morozov$^{a,b,c,\dag}$}
      {\bf A. Popolitov$^{a,b,c,\ddag}$}
      \vspace{1cm}}
}}

\maketitle
\vspace{-6cm}

\begin{center}
	\hfill ITEP/TH-09/21\\
	\hfill IITP/TH-06/21\\
	\hfill MIPT/TH-05/21
\end{center}

\vspace{2.2cm}

\begin{center}
  $^a$ {\small {\it Institute for Theoretical and Experimental Physics, Moscow 117218, Russia}}\\
  $^b$ {\small {\it Institute for Information Transmission Problems, Moscow 127994, Russia}}\\
  $^c$ {\small {\it Moscow Institute of Physics and Technology, Dolgoprudny 141701, Russia }} \\
  \vspace{0.25cm}
  $*$ {\small {\it anokhina@itep.ru}} $\dag$ {\small {\it morozov.itep@mail.ru}}  $\ddag$ {\small {\it popolit@gmail.com}}
\end{center}

\vspace{0cm}

\begin{abstract}
	We compute explicitly the Khovanov polynomials (using the computer program from katlas.org) for the two
	simplest families of the satellite knots, which are the twisted Whitehead doubles and the two-strand cables.
	We find that a quantum group decomposition for the HOMFLY polynomial of a satellite knot
	can be extended to the Khovanov polynomial, whose quantum group properties are not manifest. Namely, the Khovanov polynomial of a twisted Whitehead
	double or two-strand cable (the two simplest satellite families) can be presented as a naively
	deformed linear combination of the pattern and companion invariants.
	For a given companion, the satellite polynomial ``smoothly'' depends on the pattern but for the ``jump'' at one critical point defined by the $s$-invariant of the companion knot. 
A similar phenomenon is known for the knot Floer homology and $\tau$-invariant for the same kind of satellites 
	
\end{abstract}

\tableofcontents

\section{Introduction\label{sec:int}}



\paragraph{Why satellites?}
One can image a satellite knot as a knot inside (i.e., non-tivially embedded into) a solid torus tied in the form of another (\textit{companion}) knot \cite{KnotBook}. In this paper, we study the Khovanov polynomials of the satellite knots. We see at least two reasons to do this.

First, the satellite knots are interesting in themselves. Being rare among relatively simple prime knots, the satellite knots can actually predominate in the entire knot space \cite{MalBelSat,MalSat}. This is already a reason to develop efficient calculus for associated knot polynomials.
Apart from that, the Khovanov polynomial of a satellite is actually the polynomial of a knot in a solid torus. Thus we get another approach to the knots in the simplest handled body, for which these invariants invariants are yet little studied \cite{KhHand}.

Second, the satellite knots play a major role in definition the coloured knot polynomials, both of the Jones/HOMFLY \cite{AMcab} and Khovanov/-Rozansky \cite{KhoCol,HedCol} types. 
However, a relation between the satellite polynomials and ``irreducible'' coloured polynomials is much more involved in the latter case \cite{MacTurCol,BelWehCol,CapCol,ItoCol}. In fact, the corresponding polynomials are well studied only in extreme cases, such as higher colour \cite{RozCol} and infinite braid \cite{WilCol} limits. Only very simplified versions of these invariants were studied for more or less general knots \cite{MacTurCol,RosWedCol}.
Although several ``combinatorial'' definitions of the coloured Khovanov-Rozansky homology were constructed \cite{QueRosCol, RobWagcol},  none of them (as far as we know) is still realised as an effective homology calculator. In contrast, there are such calculators for plain Khovanov and for particular cases of Khovanov-Rozansky homologies and polynomials \cite{LukHomCal}. Hence one can try to extract a candidate for a ``coloured Khovanov-type invariant'' of a knot from the explicitly computed Khovanov polynomials of its satellites.

Another part of the story is the knot Floer homology, which is believed to be an other (``dual'' to Khovanov) ``reduction'' of the superpolynomial \cite{RobWag, SlepNovAl}. The Floer homologies were studied both for the Whitehead doubles \cite{HedWh} and cables \cite{HedCab1, HedCab2}. Hence the next step is to ``embed'' the Floer and Khovanov results into a new knowledge about the knot superpolynomials.

\begin{figure}
	$$
	\arraycolsep=1cm
	\begin{array}{cc}
	\includegraphics[width=4cm]{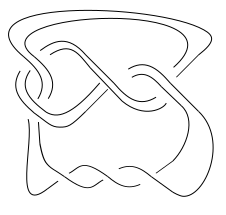}&
	\includegraphics[width=4cm]{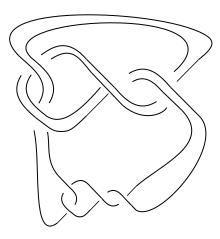}\\
	\mathrm{I.}\ \Sm_{\Tor_{3}}^{\Tw_{-2}}&\mathrm{II.}\ \Sm_{\Tw_{-2}}^{\Tw_{-2}}\\
	\end{array}
	$$
	\caption{Torus and twist satellites of the figure-eight knot\label{fig:sat}}
\end{figure}

\paragraph{Particular problem we study.}
In \cite{MorTwSat}, a simple procedure was suggested to build the
double-graded HOMFLY-PT polynomials \cite{HOMFLY,PT,Kaul,AMcab}
for satellite knots, especially for the Whitehead doubles (fig.\ref{fig:sat}, on the right).
It is based on the $\Rm$-matrix formalism for HOMFLY polynomials \cite{ReshTur,MorSm,AMcab,TabArb}
and reduces to calculating a peculiar ``lock'' block and
the uniform adjoint polynomials \cite{MirMorUn}.
This is a powerful approach, straightforwardly generalisable to other knot families
and to the coloured polynomials. Yet, an extension to the triply graded superpolynomials \cite{DunGukRas,KhR}, or to quadruply graded  hyperpolynomials \cite{Gor4gr, Arth4gr} is still a big puzzle. An attempt to find one was already made in \cite{MorTwSat}, but at that time there was no practical way to check and improve the suggested answers.
Now we can do this, at least partly, by comparing conjectures with the explicitly calculated
Khovanov polynomials. This is the goal of the present study. 

\begin{figure}
	$$
	\arraycolsep=1cm
	\begin{array}{cc}
	\includegraphics[width=2cm]{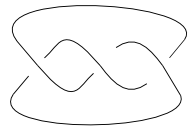}&
	\includegraphics[width=2cm]{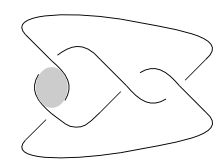}\\
	\mathrm{I.}\ \Tor_{3}&\mathrm{II.}\ \Tw_{-2}\\
	\end{array}
	$$
\caption{Two-stand torus and twist knots\label{fig:tortw}}
\end{figure}

\paragraph{Summary of results.}
Below we summarize our main observations and conclusions of the explicitly computed Khovanov polynomials of the simplest satellite knots and knot families by means of the (slightly improved) program from \cite{katlas}.

\paragraph{$\dms$ Positive pattern-companion decomposition (PPCD).} Our main conjecture is an extension of the pattern-companion decomposition from \cite{MorTwSat} to the Khovanov polynomials. Namely, the Khovanov polynomial of a twisted Whitehead
double or two-strand cable (the two simplest satellite families) can be presented as a naively
deformed linear combination of the positive polynomial invariants of the pattern and companion,
\be
\Hm^{\Sm_{\widetilde{\Km}}^{\Km}}(A,q)=f(A,q)\Hm_{\widetilde{\Km}}(A,q)+g(A,q)\Hfr_{\Km}(A,q)\ \longrightarrow\ 
\Kh^{\Sm_{\widetilde{\Km}}^{\Km}}(q,t)=F(q,t)\Kh_{\widetilde{\Km}}(q,t)+G(q,t)\Kfr_{\Km}(q,t).
\ee 
In sec.\ref{sec:PPCD} we give explicit formulas (\ref{ppcd-tor-sat}) and (\ref{ppcd-tw-sat}) for the torus and twist cases.

\textbf{$\dm$ The pattern invariant in $\Kh_{\widetilde{\Km}}(q,t)$} is the Khovanov polynomial of the pattern-defining twisted or torus knot, respectively. 

\textbf{$\dm$ The companion invariant in $\Kfr_{\Km}(q,t)$} is a positive polynomial, which is likely to be a ``tail'' of the adjoint Khovanov that survives in the satellite polynomial (unlike the true adjoint Khovanov itself). 

We determine both these quantities explicitly for the simplest companion knots and knot families by writing and solving the difference ``evolution'' equations for the computed satellite knot polynomials, treating the half-twist number in the pattern as a variable. 

\paragraph{$\dms$ Critical point of evolution flow and the $s$-invariant.} For a given companion, the satellite polynomial ``smoothly'' depends on the pattern but for a single ``$\Theta$-jump'' in $\Kh_{\widetilde{\Km}}(q,t)$ (see sec.\ref{sec:patt-pol}). This jump happens when an invariant combination of the pattern's hail-twist number and the companion's writhe equals three times Rasmussen $s$-invariant, which coincides with the knot signature for all considered knots (see sec.\ref{sec:sat}). 

A similar phenomenon is known for the knot Floer homology and Ozsv\'ath-Szab\'o $\tau$-invariant\footnote{The two invariants are related as $t=s/2$ for all knots we take as \textit{companions}, but not in general. Counterexamples are the Whitehead doubles of the two-strand torus knots \cite{HedOrd}.}  for the same kind of satellites 
\cite{HedWh, HedCab1,HedCab2}.  

%

\section{The knots we study and the questions we ask\label{sec:knoque}}
In this section we describe the simplest families of satellites knots and their relation to the coloured knot polynomials. We see at least two reasons to use the HOMFLY polynomial as a knot identifier. First, many statements look simpler in terms of knot invariants than in purely geometric terms (far from being a complete invariant, the HOMFLY polynomial is good enough for our purposes). Second, it is the properties of the HOMFLY polynomials that we wish to extend to the Khovanov case.

\

The standard definition of a \textbf{satellite knot} $\Sm_{\widetilde{\Km}}^{\Km}$ (see, e.g., Definition~2.8 of  \cite{KnotBook}) includes the three following items.
\begin{enumerate}
	\item{A knot $\widetilde{\Km}$ is non-trivially embedded into an \textit{unknotted} solid torus; i.e., the $\widetilde{\Km}$ neither lie in a ball inside the solid torus, nor is homotopic to the central core curve of the solid torus.}
	
	\item{The solid torus together with the knot $\widetilde{\Km}$ is subjected to a homomorphism and becomes a tubular neighbourhood of another non-trivial knot $\Km$.}
	\item{The longitude of the original solid torus is mapped into the longitude of its image.}
\end{enumerate}
Recall that a longitude of a solid torus is a simple closed curve that (i) has zero linking number with the central curve of solid torus \textit{or}, equivalently, (ii) has zero linking number with its image under a parallel shift along the torus and non-contractible in the solid torus.

The knot $\Km$ is called a \textit{companion}, and the pair $(\Km,\widetilde{\Km})$ is called a \textit{pattern} of the satellite. 

The third item essentially completes the definition. Otherwise the pattern would be defined up to a Dehn twist of the solid torus. In particular, each such twist affects a satellite's planar diagram like one in fig.\ref{fig:sat} by changing the number of crossings between the two strands of the ``thickened'' companion. The resulting diagram represents a topologically distinct satellite (see sec.\ref{sec:sat}).

The knot families that we consider naturally include the degenerate cases with the unknot as $\Km$ or $\widetilde{\Km}$. We still consider these cases, although such knots do not fit the definition of a satellite (otherwise \textit{any} knot would be a satellite of the unknot or of itself).

\subsection{Coloured decompositions for the simplest satellites}
Many colored knot polynomials (the symmetric and adjoint HOMFLY are among them) can be computed (or even defined) as linear combinations of plain polynomials for the knot cables~\cite{AMcab} and, more generally, for the knot satellites~\cite{KnotBook}. The simplest family, the torus satellites $\Sm_{\Tor_k}^{\Km}$, looks like multiply twisted ribbons tied in the (companion) knot $\Km$ (fig.\ref{fig:sat}.I). The closure of the ribbon is then a two-strand torus knot (fig.\ref{fig:tortw}.I). The \textit{unreduced} plain HOMFLY polynomials of these satellites are linear combinations of the unreduced coloured (namely, antisymmetric and symmetric) polynomials of the knot $\Km$~\cite{AMcab},
\be
H^{\Sm_{\Tor_k}^{\Km}}=-(Aq)^{-k}H^{\Km}_{\asm}+(A/q)^{-k}H^{\Km}_{\sym}.\label{tor-cab-top_u}
\ee
The power $k$ is related to the number of half-twist of the ribbon as we specify below.

If the border of the ribbon is linked with itself to form a twist knot (fig.\ref{fig:tortw}.II), one obtains the next to simplest family, the twisted satellites $\Sm_{\Tw_k}^{\Km}$ (Whitehead doubles) of the knot $\Km$ (fig.\ref{fig:sat}.II). The unreduced HOMFLY of these satellites are expressed via another pair of the unreduced coloured polynomials (namely, the trivial and adjoint ones) as
\be
H^{\Sm_{\Tw_k}^{\Km}}=\tau_{\trv}H^{\Km}_{\trv}+A^{-k-1}H^{\Km}_{\adj}.\label{tw-cab-top-u}
\ee
The rational form-factor $\tau_{\trv}=A^{-2}\Big(q^2+q^{-2}-A-A^{-1}\Big)$ is associated with the ``lock-down'' pair of crossings (adjacent to the gray area in fig.\ref{fig:tortw}.II), which is new for the twisted satellites compared to the torus family.
Below we substitute $H^{\Km}_{\trv}=1$, which holds for any $\Km$. 

We will mostly work with the \textit{reduced} polynomials $\Hm$, which are ratios of the unreduced polynomials of the given knot and the unknot (in the given representation), i.e.,  $\Hm^{\Km}_Q=H^{\Km}_Q/H^{\emptyset}_Q$. The analogues of decompositions (\ref{tor-cab-top_u},\ref{tw-cab-top-u}) for the reduced polynomials are
\be
\Hm^{\Sm_{\Tor_k}^{\Km}}=A^{-k}\Big(-q^{-k}\frac{\{Aq^{-1}\}}{\{q^2\}}\Hm^{\Km}_{\asm}+q^{k}\frac{\{Aq\}}{\{q^2\}}\Hm^{\Km}_{\sym}\Big),
\label{tor-cab-top}
\ee
\be
\Hm^{\Sm_{\Tw_k}^{\Km}}=A^{-2}\Big(1+\frac{A\{q\}^2}{\{A\}}+A^{-k+1}\frac{\{Aq\}\{A/q\}}{\{A\}}\Hm^{\Km}_{\adj}\Big).\label{tw-cab-top}
\ee

\subsection{The pattern decomposition\label{sec:unsat}}
Following \cite{MorTwSat}, one can identically rewrite (\ref{tw-cab-top}) and (\ref{tor-cab-top}) in form of \textit{pattern decompositions}. Namely,
\be
\Hm^{\Sm_{\Tw_k}^{\Km}}=\Hm^{\Tw_{k+\gamma}}+A^{-k-\gamma+1}\{Aq\}\{A/q\}\Hfr^{\Km}_{\adj},\label{tw-patt-v}
\ee
and
\be
\Hm^{\Sm_{\Tor_k}^{\Km}}=\left(Aq\right)^{\delta}\Hm^{\Tor_{k+\delta}}-(Aq)^{k+\delta}\{Aq^{-1}\}\Hfr^{\Km}_{\asm}+(A/q)^{k+\delta}\{Aq\}\Hfr^{\Km}_{\sym},
\label{tor-patt-v}
\ee
where
\be
\Hfr^{\Km}_{\asm}\!=\!\frac{\left(Aq\right)^{\delta}}{\{q^2\}}\Big(\Hm^{\Km}_{\asm}-1\Big),\  \Hfr^{\Km}_{\sym}\!=\!\frac{(A/q)^{-\delta}}{\{q^2\}}\Big(\Hm^{\Km}_{\sym}-q^{2\delta}\Big), 
\ \text{and}\ \Hfr^{\Km}_{\adj}\!=\!\frac{A^{\gamma}}{\{A\}}\Big(\Hm^{\Km}_{\adj}-A^{-\gamma}\Big)
\ee
are Lauren polynomials in $A$ and $q$ with integer coefficients.
One can chose \textit{any} integers for $\gamma$ and $\delta$, but they must be knot invariants if one wishes the $\Hfr$ to be a knot invariant too. Some values $\delta=\gamma\equiv\Sh_{\Km}$ given by (\ref{ShKn}) prove to be distinguished. Namely, (\ref{tor-patt-v}) then can be extended to the Khovanov case so that the coefficients are naturally ``deformed'', and $\Hfr_{\sym}$ and $\Hfr_{\adj}$ (which coincide for $A=q^2$) are substituted with positive polynomials $\Kfr$ (while $\Hfr_{\asm}$ vanishes for $A=q^2$). The distinguished value $\Sh_{\Km}$ depends on the companion knot and equals to the power of $q^2$, which acquires the factor of $(-t)$ at the critical point of the evolution flow (see sec.\ref{sec:int}). We redefine the satellite label $k\rightarrow k+\Sh_k$ in the Khovanov-related formulae to simplify them.

\subsection{How to get the satellite class from a knot diagram\label{sec:sat}}
\paragraph{The first Reidemeister of the companion and the full twist in the pattern.}
The parameter $k$ of the satellite defined by (\ref{tor-cab-top},\ref{tw-cab-top}) is a topological invariant, since it enters in the relations of the topologically invariant quantities. On the other hand, a satellite is often presented with its the planar diagram, e.g., with one in fig.\ref{fig:sat}. Then $k$ must be expressed via the number of the half-twists in the pattern, $w$, and the writhe number of the companion knot $\Km$, $\nu$. Both these quantities change under the same continuous transformation of the knotted tube, which deletes a crossing with adjacent contractible loop on the knot diagram (the RI transformation \cite{KnotBook}) and causes a full twist of the tube. The attached two-strand braid gains then two more half-twists of the same orientating as the deleted crossing\footnote{The $m$-strand case is discussed in App.\ref{app:FT}} ~\cite{KnotBook}.
Hence, $2\nu+w=inv$. Only one linear combination of $\nu$ an $w$ must be a topological invariant, since the both numbers would be invariants otherwise. Hence, $k=2\nu+w+c$, for some constant $c$.  The form of the Khovanov polynomial as function of the pattern dictates the natural chose of $c=\Sh_{\Km}$ given by (\ref{ShKn}) (see sec.\ref{sec:PPCD} for further details). 

\paragraph{Satellite class in the cabling formula.}
The invariance of $k=2\nu+w$ also follows from computation of the HOMFLY polynomial using the knot diagram. E.g., one can take the cabling formulae for torus satellites \cite{AMcab}. These formulae are originally applied to the renormalised $\check{H}$ polynomials and read 
\be
\check{H}^{\Sm_{\Tor_k}^{\Km}}=-q^{w}\check{H}^{\Km}_{\asm}+q^{-w}\check{H}^{\Km}_{\sym},\label{tor-cab-vert-u}
\ee
Each of the knot polynomials $H$, $H_{\asm}$, $H_{\sym}$ differs from $\check{H}$, $\check{H}_{\asm}$, $\check{H}_{\sym}$, respectively, by the known factor whose power depends on the knot diagram\footnote{The ratio of the topological and the vertical framings \cite{AMcab},~sec.5.}. The explicit form of the factors is such that (\ref{tor-cab-vert-u}) becomes 
\be
A^{-4\nu-w}H^{\Sm_{\Tor_k}^{\Km}}&=&-q^{w}A^{-2\nu}q^{2\nu}H^{\Km}_{\asm}+q^{-w}A^{-2\nu}q^{-2\nu}H^{\Km}_{\sym},\nn\\
H^{\Sm_{\Tor_k}^{\Km}}&=&-(Aq)^{2\nu+w}H^{\Km}_{\asm}+(A/q)^{2\nu+w}H^{\Km}_{\sym},
\label{tor-cab-top-u}\ee
and the comparison with (\ref{tor-cab-vert-u}) gives $k=2\nu+w$. Again, one can redefine the topological class by adding an arbitrary constant, $k\to k+c$.

\paragraph{What is $k$ on the standard diagram?}
A satellite knot is commonly presented with a standard planar diagram, where a ``thick'' companion is attached to an ``extra'' element. In our cases this element has a form of the two-strand or Whitehead tangle (as in fig.\ref{fig:sat}). Then it \textit{seems} natural to use the torus or twist knot that is the closure of this tangle (fig.\ref{fig:tortw}) to determine the pattern. Yet this is generally wrong, because (as follows from the above discussion) the ``extra'' tangle has a sense only together with the companion knot, and moreover with the framing of its ``thickened'' version. 

However, the half-twist number in the attached tangle does define the pattern for a given framed companion. We introduce for \textit{each} companion its own pattern variable $k=w+2\nu+\Sh_{\Km}$ with the respective last term from (\ref{ShKn}).
In fig.\ref{fig:sat}, $\nu=0$ and $\Sh=0$, and hence $k=w$ (but this is not so generally).
\subsection{Summary of Jones formulae\label{sec:Jones}}
Now we address to the case which is currently the only definite common point of the HOMFLY and Khovanov cases. This is the case of Jones polynomial, which is obtained for $A=q^2$ from the former and for $t=-1$ from the latter ones. Explicitly, the Jones polynomial of the satellites we consider has the form
\be
\arraycolsep=0.5mm
\begin{array}{ccccccc}
	J^{\Sm_{\Tor_k}^{\Km}}(q)&=&q^{-3k+1}\Big(-\cfrac{1}{1+q^2}&+&q^{2k-2}\cfrac{1-q^6}{1-q^4}&J^{\Km}_{\adj}(q)\Big),\\
	J^{\Sm_{\Tw_k}^{\Km}}(q)&=&q^{-4}\Big(\phantom{-}\cfrac{1+q^4}{1+q^2}&-&q^{-2k\phantom{-2}}\cfrac{1-q^6}{1+q^2}&J^{\Km}_{\adj}(q)\Big).\label{Jcab}
\end{array}
\ee
The antisymmetric Jones as well as the trivial Jones identically equals one. Both kinds of the satellite polynomials are now expressed via the same coloured Jones, which is the descendent of both the adjoint and symmetric HOMFLY and explicitly equals 
\be
\arraycolsep=1mm
\begin{array}{ccrccccc} 
	J^{\Tor_n}_{\adj}(q)&=&q^{-8n}\!\cdot\!\hspace{-5mm}&\bigg(\cfrac{q^2(q^2-1)}{q^6-1}&-&q^{2n}&+&\cfrac{q^{10}-1}{q^2(q^6-1)}q^{6n}\bigg),\\[8mm]
	J^{\Tw_n}_{\adj}(q)&=&&\cfrac{(q^{18}-1)(q^2-1)}{q^{12}(q^6-1)^2}&+&q^{-12}(1+q^6)(1-q^4)q^{-2n}&+&\cfrac{(q^{10}-1)(q^4-1)(q^2-1)}{q^{10}(q^6-1)}q^{-6n}
	\label{Jcol}
\end{array}
\ee
We present these expressions for the sake of reference. Namely, one can verify that (\ref{ppcd-tw-sat},\ref{ppcd-tor-sat}) are reduced to (\ref{Jcab}), and the polynomials (\ref{FAK}) explicitly given in sec.\ref{sec:AAK-tor},\ref{sec:AAK-tw} are reduced to (\ref{Jcol}) for $t=-1$.


\paragraph{Disclaimer.} Note that $n$ in (\ref{Jcol}) must be odd for torus knots and even for twist knots (see sec.\ref{app:coltor},\ref{app:coltw}). 

If one substitutes $-q^{-2n}$ with  $(-q^2)^{-n}$ in the torus formula, it yields the correct adjoint polynomials both for torus knots with odd $n$ and torus links with even $n$. The twist formula with $(-q^2)^{-n}$ instead of $q^{-2n}$ also gives a polynomial for any $n$, and these polynomials satisfy $J^{\Tw_n}_{\adj}(q)=q^{16}J^{\Tw_{-n+1}}_{\adj}(q^{-1})$. I.e., the formula still gives the true adjoint polynomials of the twist knots for even $n$, while the one with the odd $n$ now gives the polynomials of the mirror knots up to the extra factor of $q^{16}$. This property seems to be accidental and does not survive neither in HOMFLY, nor in Khovanov cases. In other words, there is no ``analytic'' formula for twist knots with any integer $n$ as the half-twist number (see sec.\ref{app:coltw}).

\section{Khovanov-evolution formulae for satellite families\label{sec:Khsat}}
Now we return to the Khovanov polynomials and formulate the PPCD conjecture form sec.\ref{sec:int} as a precise statement.
\subsection{Preliminaries}
By definition, a positive polynomial is a Laurent polynomial in $q$ and $t$ with non-negative coefficients. Moreover, in our cases these coefficients are integers by construction. Throughout the text \be\{x\}=x-x^{-1}.\label{figx}\ee
The jumps in the Khovanov polynomials at critical points of the evolution flow are captured by the sign and step functions, 
\be
\sgn_x=\left\{\begin{array}{cl}-1,&x<0\\0,&x=0\\1,&x>0\end{array}\right.,\ \ \
\Theta_x=\left\{\begin{array}{cl}0,&x\le0\\1,&x> 0\end{array}\right..
\ee

\subsection{Positive pattern-companion decompositions\label{sec:PPCD}}
$\star$ The reduced Khovanov polynomial of a twisted satellite (Whitehead double) of a knot $\Km$ has the form ($k$ in $\Tw_k$ is \textit{even}, see \ref{app:coltw})
\be
\Kh^{\Sm_{\Tw_k}^{\Km}}=\Kh^{\Tw_{k}}(q,t)+\left(q^2t\right)^{-k-2}\left(1+q^6t^3\right)\left(1+q^2t\right){\Kfr}^{\Km}(q,t).
\label{ppcd-tw-sat}
\ee
$\star$ The reduced Khovanov polynomial of a two-strand torus satellite (two-strand cable) of a knot $\Km$ has the form
($k$ in $\Tor_k$ to be \textit{odd}, see \ref{app:coltor})
\be
\Kh^{\Sm_{\Tor_k}^{\Km}}=\left(q^3t\right)^{-\Sh_{\Km}}\!\cdot\!\Big(
\Kh^{\Tor_{k}}(q,t)+t^{-1}q^{k-1}\left(1+q^6t^3\right){\Kfr}^{\Km}(q,t)\Big).
\label{ppcd-tor-sat}
\ee
$\star$ The unreduced Khovanov polynomials in the same case can be presented as
\be
{}_u\Kh^{\Sm_{\Tor_k}^{\Km}}=\left(q^3t\right)^{-\Sh_{\Km}}\!\cdot\!\Big(
{}_u\Kh^{\Tor_{k}}(q,t)+q^{k}{_u}{\hat\Kfr}^{\Km}(q,t)-
q^{k}\left(1+q^{-2\sgn_n}(-t)^{\Theta_k-\Theta_n}\right)\Big).
\label{ppcd-u-tor-sat}
\ee
The above formulae contain the following quantities,
\be
\arraycolsep=1mm
\begin{array}{ccp{15cm}}
	\dms&k&is an integer-valued topological invariant that defines the pattern. \\
	&\dm&The $k$-satellites of the unknot are $\Sm^{0_1}_{\Tw_k}=\Tw_k$ and $\Sm^{0_1}_{\Tor_k}=\Tor_k$;\\[2mm]
	&\dm&We explain how to get the $k$ from a planar diagram of the satellite in sec.\ref{sec:sat}.\\[2mm]
	\dms&\Kfr^{\Km}& is a positive polynomial invariant of the companion $\Km$. \\
	&\dm& The reduced invariant vanishes for the unknot, i.e., $\Kfr^{0_1}=0$.\\
	&\dm& \textit{The same} $\Kfr^{\Km}$ enters all formulae. \\[2mm]
	\dms&\Sh_{\Km}& is an integer-valued invariant of the companion $\Km$; \\ 
	&\dm&in all studied cases, given in (\ref{ShKn}), $\Sh_{\Km}=-3s_{\Km}$, where $s$ is both the knot signature and the Rasmussen invariant.  
\end{array}
\nn\ee
Below we recall the values of $\Sh^{\Km}$ for the studied knots (up to the knot $8_9$ in \cite{katlas}, in sec.\ref{sec:AAK-rolf}) and knot families (the two-strand torus knots, in sec.\ref{sec:AAK-tor}, and the twist knots, in sec.\ref{sec:AAK-tw}).
\be
\arraycolsep=0.5mm
\begin{array}{|c|c|c|c|c|c|c|c|c|c|c|c|c|c|c|c|c|c|c|c|c|c|c|c|}
	\hline\rule{0pt}{4mm}
	\Km&3_1&4_1&5_1&5_2&6_1&6_2&6_3&7_1&7_2&7_3&7_4&7_5&7_6&7_7&8_1
	&8_2&8_3&8_4&8_5&8_6&8_7&8_8&8_9\\
	\hline
	\simeq&\Tor_{3}&&\Tor_5&&&&&\Tor_7&&&&&&&&&&&&&&&\\
	\hline
	\simeq&\rule{0pt}{3mm}\overline{\rule{0pt}{3mm}\Tw}_{2}&\Tw_{-2}&&\Tw_3&\Tw_{-4}&&&&\Tw_5&&&&&&\Tw_{-6}&&&&&&&&\\
	\hline
	\Sh_{\Km}&6&0&12&6&0&6&0&18&6&-12&-6&12&6&0&0&12&0&6&-12&6&-6&0&0\\
	\hline\hline
	\multicolumn{6}{|c|}{\rule{0pt}{4mm}\Sh_{\Tor_{n<0}}=3n+3}&\multicolumn{6}{c|}{\Sh_{\Tor_{n>0}}=3n-3}&
	\multicolumn{6}{c|}{\Sh_{\Tw_{n\le0}}=0}&\multicolumn{6}{c|}{\Sh_{\Tw_{n>0}}=6}\\[0.5mm]
	\hline
\end{array}
\label{ShKn}
\ee

\subsection{False adjoint Khovanov polynomials}
We call the invariant $\Kfr^{\Km}$ an \textbf{asymptotic adjoint Khovanov}/\textbf{AAK} polynomial. It is \emph{almost} the adjoint/symmetric Khovanov polynomial. Namely, the polynomial
\be\widehat{\Kh}^{\Km}_k(q,t)=
(-t)^{-1}\left(q^2t\right)^{-\Sh_{\Km}}\Big((-t)^{\Theta_k}-\left(1-q^4t^2\right){\Kfr}^{\Km}(q,t)\Big)\label{FAK}
\ee
stands for the symmetric polynomial in (\ref{ppcd-tw-sat},\ref{ppcd-tor-sat}) when one \textit{naively} compares the decompositions of the satellite polynomials over the coloured ones in the Khovanov and in the HOMFLY cases. In the particular case $t=-1$ one gets the first coloured Jones, 
\be\widehat{\Kh}^{\Km}_k(q,t=-1)=J_2(q)\label{J-Kh-col}\ee
The $\widehat{\Kh}^{\Km}_k$  is \textit{not} a positive polynomial, but $\widehat{\Kh}^{\Km}_k/(1-q^4t^2)$ is expanded into a positive Laurent series in $t^{\sgn_k}$. The unreduced counterpart equals
\be
{_u}\widehat{\Kh}^{\Km}_k(q,t)=\left(q^2t\right)^{-\Sh_{\Km}}\Big(q^{-2}(-t)^{\Theta_k}+{{_u}\Kfr}^{\Km}(q,t)\Big)
\label{TBAKu}\ee
and satisfies
\be{_u}\widehat{\Kh}^{\Km}_k(q,t=-1)=\textstyle{\frac{1+q^2+q^4}{q^2}}J_2(q).\label{Ju-Khu-col}\ee
The non-positivity of (\ref{FAK}) can imply that the satellite polynomial, which is the Poincare polynomial of the corresponding Khovanov complex, does not contain \textit{all} ``trivial-representation'' generators (and hence becomes sign-indefinite when we naively subtract them all).

\section{Explicit form of the pattern and companion invariants\label{sec:AAK}}
Now we give an explicit form of the two basic ingredients in PPCD formulae from sec.\ref{sec:PPCD}. First, we describe the pattern invariant in sec.\ref{sec:patt-pol}. In fact, we recall the well known Khovanov polynomials of the two-strand torus and twist knots, bringing the expressions to a proper form. In next sec.\ref{sec:AAK-rolf},\ref{sec:AAK-tor},\ref{sec:AAK-tw}, we present the explicit expressions for the companion invariants extracted from PPCD. We call them an asymptotic adjoint Khovanov (AAK) polynomials, since we believe them to be ``tails'' of the true adjoint Khovanov polynomials that survive in the satellite polynomials. Although the exact relation of the two quantities is not yet established, the AAK may be interesting in itself, as an explicitly defined and readily computable Khovanov-like knot invariant. 

\subsection{The pattern-defining polynomial\label{sec:patt-pol}}
By definition, the $k$ in our $\Sm_{\Tw_k}$ and $\Sm_{\Tor_k}$, as well as in (\ref{ppcd-tw-sat},\ref{ppcd-tor-sat},\ref{ppcd-u-tor-sat}) is a satellite invariant independent of the companion knot. Hence $k$ and thus the knot $\Tw_k$ or $\Tor_k$ defines the pattern of the satellite (see sec.\ref{sec:sat}). The first term in the PPCD decomposition is just the respective Khovanov polynomial.  
Namely, the reduced polynomial for the twist knot is 
\be
\Kh^{\Tw_k}&=&-(-t)^{\Theta_{k}}\left(q^2t\right)^{-k-2}\!\cdot\!
\Bigg(1+q^2t\!\cdot\!\frac{\left(1+q^4t^2\right)\left(1-(q^2t)^{k-1}\right)}{1-q^2t}
\Bigg)\label{Kh-tw}\\\nn
&=&
(-t)^{\Theta_{k}}\left(q^2t\right)^{-k}\!\cdot\!
\Bigg(1+\left(q^2t\right)^{k-2}\!\cdot\!\frac{\left(1+q^4t^2\right)\left(1-(q^2t)^{-k}\right)}{1-q^2t}
\Bigg)=\\
&=&
(-t)^{\Theta_k}(q^2t)^{-2}\Bigg(\cfrac{1+q^4t^2}{1-q^2t}-(q^2t)^{-k}\cfrac{1+q^6t^3}{1-q^2t}
\Bigg),
\nn\ee
while the reduced polynomial for the torus knot is 
\be
\Kh^{\Tor_k}&=&
-(-t)^{\Theta_{k}}t^{-1}q^{k-1}\Bigg(1+\left(q^2t\right)^2\!\cdot\!\frac{1-(q^2t)^{k-1}}{1-q^2t}
\Bigg)\label{Kh-tor}\\\nn&=&
(-t)^{\Theta_{k}}q^{k+1}\Bigg(1+\left(q^2t\right)^k\cdot\!\frac{1-(q^2t)^{-k-1}}{1-q^2t}
\Bigg)=\\
&=&(-t)^{\Theta_k}q(q^3t)^k\Bigg(\cfrac{1}{1-q^2t}-(q^2t)^{-k-1}\cfrac{1+q^6t^3}{1-q^4t^2}\Bigg),
\nn\ee
and the unreduced polynomial for the torus knot is
\be
{{}_u\Kh}^{\Tor_k}&=&
q^k\!\cdot\!\Bigg\{1-(-t)^{\Theta_{k}}\Bigg(\frac{1}{q^2t}+q^2t\!\cdot\!\frac{\left(1+q^4t\right)\left(1-(q^2t)^{k-1}\right)}{1-q^4t^2}\Bigg)\Bigg\}=\\&=&
q^k\!\cdot\!\Bigg\{1+(-t)^{\Theta_{k}}\Bigg(q^2+\left(q^2t\right)^k\!\cdot\!\frac{\left(1+q^4t\right)\left(1-(q^2t)^{-k-1}\right)}{1-q^4t^2}\Bigg)\Bigg\}.\nn\ee
We give several equivalent expressions in all cases.
The first and the second expressions for each of the polynomials are manifestly positive for $n>0$, and $n<0$, respectively. The third expression for each of the reduced polynomials is useful to rewrite our PPCD formulae (\ref{ppcd-tor-sat},\ref{ppcd-tw-sat}) like (\ref{tor-cab-top},\ref{tw-cab-top}) and (\ref{Jcab}).

\subsection{AAK for the simplest prime knots\label{sec:AAK-rolf}}
We studied decompositions (\ref{ppcd-tw-sat}) for the simplest prime knots up to $8_9$. Below we present the explicit expressions for the corresponding AAK of the prime knots with no more than 6 crossings.
\be
\begin{array}{cp{16cm}}
\Km&${\Kfr}^{\Km}(q,t)$\\
3_1&$1/q^2+1/(q^{10}t^5)+1/(q^8t^4)+t^2+q^4t^4$\\[2mm]
4_1&$1+1/q^2+1/(q^{12}t^7)+1/(q^{10}t^6)+1/(q^6t^3)+2/(q^4t^2)+1/(q^2t)+1/(q^4t)+2t+q^2t^2+q^6t^5+q^8t^6$\\[2mm]
5_1&$1/q^2+1/(q^{14}t^7)+1/(q^{12}t^6)+1/(q^{10}t^5)+1/(q^8t^4)+1/(q^8t^3)+1/(q^6t^2)+t/q^2+2t^2+q^4t^4+q^6t^6+q^8t^8+q^{12}t^{10}$\\[2mm]
5_2&$2/q^2+1/q^4+1/(q^{22}t^{13})+1/(q^{20}t^{12})+1/(q^{16}t^9)+2/(q^{14}t^8)+1/(q^{12}t^7)+1/(q^{14}t^7)+2/(q^{12}t^6)+2/(q^{10}t^5)+2/(q^8t^4)+1/(q^6t^3)+2/(q^6t^2)+3/(q^4t)+t+2t^2+q^2t^3+q^4t^4$
\\[2mm]
6_1&$1+2/q^2+1/(q^8t^3)+q^6t^5+q^2t^3+q^4t^4+2q^2t^2+1/(q^{24}t^{15})+1/(q^{22}t^{14})+1/(q^{18}t^{11})+2/(q^{16}t^{10})+1/(q^{14}t^9)+3/(q^4t)+1/(q^2t)+4/(q^6t^3)+1/(q^6t^2)+4/(q^4t^2)+q^8t^6+2/(q^{10}t^6)+1/(q^8t^5)+1/(q^{16}t^9)+2/(q^{14}t^8)+2/(q^{12}t^7)+1/(q^{10}t^5)+3/(q^8t^4)+3t$\\[2mm]
6_2&$1+5/q^2+2q^{10}t^8+3q^8t^7+2q^6t^5+q^6t^6+q^4t^5+4q^2t^3+5q^4t^4+2q^2t^2+q^{16}t^{12}+q^{14}t^{11}+5/(q^4t)+1/(q^2t)+4/(q^6t^3)+2/(q^6t^2)+3/(q^4t^2)+2q^8t^6+1/(q^{10}t^6)+1/(q^{12}t^6)+1/(q^{16}t^9)+2/(q^{14}t^8)+2/(q^{12}t^7)+3/(q^{10}t^5)+4/(q^8t^4)+2t^2+4t$\\[2mm]
6_3&$2+t^2+6q^2t^2+5q^6t^5+3q^8t^6+8/(q^4t^2)+q^2t^3+1/(q^{14}t^8)+2/(q^{16}t^{10})+q^{10}t^8+q^{14}t^{10}+q^{10}t^7+q^6t^4+1/(q^6t^4)+4q^4t^3+1/(q^4t^3)+2q^{12}t^9+5/(q^{10}t^6)+4/(q^8t^4)+3/(q^{12}t^7)+1/(q^{10}t^5)+6/(q^6t^3)+5/(q^2t)+2/(q^4t)+1/(q^{18}t^{11})+1/(q^{14}t^9)+4/(q^8t^5)+8t+5/q^2+4q^4t^4$
\end{array}
\label{RolfAAK}\ee
The coloured Jones polynomials of the simplest knots are tabulated \cite{knbook} (see App.\ref{app:data}), and one can verify property (\ref{AAKJcol}). Namely,
\be\widehat{\Kh}^{\Km}_k(q,t=-1)=
\left(-q^2\right)^{-\Sh_{\Km}}\Big(1-\left(1-q^4\right){\Kfr}^{\Km}(q,t=-1)\Big)=J_2(q),\label{AAKJcol}
\ee

\subsection{AAK for the two-strand torus knots\label{sec:AAK-tor}}
We have already considered the simplest torus knots, which are the knots $T[2,n]$ with $n=3,5,7$, or the prime knots $3_1$, $5_1$, $7_1$, respectively. The AAK for the former two ones are explicitly given in (\ref{RolfAAK}). Now we study (\ref{ppcd-tor-sat},\ref{ppcd-tw-sat}), and (\ref{ppcd-u-tor-sat}) for
the two-strand torus knots $T[2,2k+1]\equiv\Tor_{2k+1}$, considering them as an entire family. The two former decompositions give the same AAK polynomials extracted from the reduced polynomials, while the latter one gives their non-trivial unreduced counterparts. Below we present explicit formulae for evolution of the AAK in $n$ (an odd integer). The formulae describe critical jumps at the unknots $n=\pm1$ and a ``smooth'' exponential dependence on $n$ in each of the domains $n<-1$ and $n>1$. 
\subsubsection{General formulae}
The unreduced (label $u$) AAK polynomials for the two-strand torus knots have the following form
\be
\Kfr^{\Tor_n}&=&t\left(q^2t\right)^{\Sh_{\Tor_n}\!\!-4n}\hspace{-5mm}\sum_{\lambda=1,q^2t,q^6t^4}\hspace{-5mm}C^{\Tor_{\sgn_n}}_{\lambda}\lambda^{n-\sgn_n},\\
{_u}\Kfr^{\Tor_n}&=&\left(q^2t\right)^{\Sh_{\Tor_n}}\Big(q^{n-\Sh_{\Tor_n}}\left(q^{-1}+q\right)+q^{-2}t^{-\Theta_n}\left(1+q^4t\right)
\left(q^2t\right)^{-4n}\hspace{-5mm}\sum_{\lambda=1,q^2t,q^6t^4}\hspace{-5mm}{_u}C^{\Tor_{\sgn_n}}_{\lambda}\lambda^{n-\sgn_n}\Big).
\label{tor-evo}
\ee
Here
\be
\sum_{\lambda=1,q^2t,q^6t^4}\hspace{-5mm}C^{\Tor_{\sgn_n}}_{\lambda}=0,\hspace{1cm}
\sum_{\lambda=1,q^2t,q^6t^4}\hspace{-5mm}{_u}C^{\Tor_{\sgn_n}}_{\lambda}=-(q^2t)^{-4\sgn_n}.
\ee
The coefficients for the reduced polynomials explicitly equal
\be
\begin{array}{|c|c|c|c|}
	\hline\rule{0pt}{4mm}
	\lambda&1&q^2t&q^6t^4\\
	\hline\rule{0pt}{7mm}
	\gC_{\lambda}^{\Tor}&\cfrac{q^2t}{(1-q^2t)(1-q^6t^4)}&-(q^6t^4)^{-\Theta_n}\cfrac{1+q^{12}t^8}{(1-q^2t)(1-q^8t^6)}&\cfrac{1+q^{14}t^9}{q^2t(1-q^6t^4)(1-q^8t^6)}\\[3mm]
	\hline\rule{0pt}{7mm}
	\ggC^{\lambda}_{\Tor}&0&-(q^6t^4)^{-\Theta_n}\cfrac{q^2t^2(1+q^6t^3)}{(1-q^4t^2)(1+q^4t^3)}&-\cfrac{q^2t^2}{1-q^8t^6}\\[3mm]
	\hline
\end{array}
\ee
The unreduced coefficients are expressed via reduced ones as (for $n<0$)
\be
\begin{array}{|c|c|c|c|}
	\hline\rule{0pt}{4mm}
	\lambda&1&q^2t&q^6t^4\\
	\hline\rule{0pt}{7mm}
	 C_{\lambda}^{\Tor_-}-{_u}C_{\lambda}^{\Tor_-}&\cfrac{q^4t^3}{1-q^6t^4}&\cfrac{1+q^2t^2}{1+q^4t^3}&
	 q^{-6}t^{-4}\left(q^{-2}\!+\!1\right)-\cfrac{1+q^6t^4}{q^6t^4(1+q^4t^3)}-\cfrac{1}{q^2t(1-q^6t^4)}\\[3mm]
	 \hline
\end{array}
\ee

\subsubsection{Reduced vs unreduced polynomials}
The unreduced Khovanov polynomials are expectedly more complicated compared to the reduced ones. In particular, one must extract an ``extra'' exceptional pair in (\ref{ppcd-u-tor-sat}), and our unreduced formulae are invalid for the unknot. The simple quantum group relations are restored for $t=-1$,
\be
\frac{\Kh_u^{\Km}(q,t=-1)}{\Kh(q,t=-1)}=\frac{\widehat{\Kh}_u^{\Km}(q,t=-1)}{\widehat{\Kh}^{\Km}(q,t=-1)}=\frac{q^4+q^2+1}{q^2}.
\ee
\subsubsection{Mirror symmetry}
By construction, the Khovanov polynomial have the mirror symmetry for all knots including the torus-torus satellites, i.e.,
\be
\Kh^{\Km}_{\Tor_{k}}(q,t)=\Kh^{\overline{\Km}}_{\Tor_{-k}}(q^{-1},t^{-1})\ \ \Rightarrow\ \ 
\Kh^{\Sm^{\Tor_n}_{\Tor_k}}(q,t)=\Kh^{\Sm^{\Tor_{-n}}_{\Tor_{-k}}}(q^{-1},t^{-1}).
\ee
Our formulae do respect this symmetry, although not explicitly. In particular, the expansion coefficients satisfy
\be C_{\lambda}^{\Tor_-}(q,t)=C_{\lambda}^{\Tor_-}(q^{-1},t^{-1})/(q^4t^3),\ \ \mbox{while}\ \ {_u}C_{\lambda}^{\Tor_+}(q,t)={_u}C_{\lambda}^{\Tor_-}(q^{-1},t^{-1}).\ee

\subsection{AAK for the twist knots\label{sec:AAK-tw}}
The simplest twist knots (with 1,2,3 half-twist) are the prime knots $3_1$, $5_2$, $7_2$ (positive half-twists), and $4_1$, $6_1$, $8_1$ (negative half-twists). The AAK for the up to 2 half-twists are explicitly given in (\ref{RolfAAK}). Below we present the explicit evolution formulae for the entire twist family $\Tw_{n}$ (an even integer $n$ is the crossing number, not to mix the half-twist number $l=\pm\frac{n-2}{2}$).  Both decompositions (\ref{ppcd-tor-sat}) and (\ref{ppcd-tw-sat}) give the same AAK.
The obtained formulae describe a ``smooth'' exponential dependence on $n$ in each of the domains $n<0$ and $n>0$, aside the critical jumps at the unknot $n=0$. 
\subsubsection{General formulae}
The AAK of the twist knots have the form
\be
\Kfr^{\Tw_n}=t\left(q^2t\right)^{\Sh_{\Tw_n}}\hspace{-5mm}\sum_{\lambda=1,q^2t,q^6t^4}\hspace{-5mm}
C^{\Tw_{\sgn_n}}_{\lambda}\lambda^{2\Theta_n-n},\ \ 
C^{\Tw_{\sgn_n}}_{\lambda}=(\lambda/t)^{-2\Theta_n}\left(\gC_{\lambda}^{\Tw}+(-t)^{-\Theta_n}\ggC_{\lambda}^{\Tw}\right),
\label{tw-evo}
\ee
with the coefficients
\be
\small{
	\arraycolsep=1mm
\begin{array}{|c|c|c|c|}
	\hline\rule{0pt}{4mm}
	\lambda&1&q^2t&q^6t^4\\
	\hline\rule{0pt}{7mm}
	\gC_{\lambda}^{\Tw}&\cfrac{1\!+\!q^6t^4\!+\!q^8t^5\!+\!q^{10}t^7}{q^{12}t^8(1\!-\!q^2t)(1\!-\!q^6t^4)}&
	\cfrac{\!-\!1}{1\!-\!q^2t}\bigg(\cfrac{(1\!+\!q^2t^2)(1\!+\!q^6t^4)}{q^{12}t^{8}}\!+\!\cfrac{(t\!\!+\!\!1)}{1\!-\!q^4t^3}\!+\!\cfrac{(t\!\!+\!\!1)^2}{q^2t^2(1\!-\!q^6t^8)}\bigg)&
	\cfrac{(1\!+\!q^2t)(1\!+\!q^8t^5)(1\!+\!q^{10}t^7)}{q^{10}t^6(1\!-\!q^6t^4)(1\!-\!q^8t^6)}
	\\[3mm]
	\hline\rule{0pt}{7mm}
	\ggC^{\lambda}_{\Tw}&\cfrac{1\!+\!q^4t(q^2t)^{\!-\!6\Theta_n}}{q^4t^2(1\!-\!q^4t^2)}&
	\cfrac{\!-\!(1\!+\!t)}{q^4t^3(1\!-\!q^4t^2)}&0\\[3mm]
	\hline
\end{array}}.
\ee
\subsubsection{Why there is no mirror symmetry?}
Because the $\Tw_+$ and $\Tw_-$ are the two distinct families, and their polynomials are not related in any simple way (see App.\ref{app:coltw}).
\begin{wrapfigure}{l}{180pt}
	$$
	\arraycolsep=0cm
	\begin{array}{cc}
	\includegraphics[width=3cm]{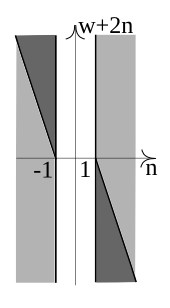}&
	\includegraphics[width=3cm]{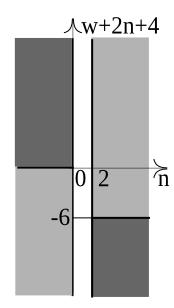}\\
	\mathrm{I.}\ \Km=\Tor_{n}&\mathrm{II.}\ \Km=\Tw_{-2}\\
	\end{array}
	$$
	\caption{Double-evolution domains for torus and twist companions.\label{fig:dbsat}}
\end{wrapfigure}
\subsection{Double-evolution phase diagrams}
Now one can combine companion-evolution formulae (\ref{tor-evo},\ref{tw-evo}) with pattern-evolution formulae (\ref{ppcd-tw-sat},\ref{ppcd-tor-sat},\ref{ppcd-u-tor-sat}) to obtain double-evolution diagrams as in \cite{DBPev,APev}. A subtle point is that the invariant pattern-evolution variable $k$ is related to the parameters of the knot diagram via a companion-dependent expression, as we discussed in sec.\ref{sec:sat}. If the companion's crossing number $n$ and the pattern's half-twist number $w$ are chosen as variables, the critical values of the pattern lie on a line determined by the relevant $\Sh_{\Km}$ from (\ref{ShKn}) for the companion family we consider. In addition, there are critical values of the companion evolution (when the companion is the unknot).  Altogether the critical lines divide the parametric plane into four evolution domains, as shown in fig.\ref{fig:dbsat}.

\section{Eigenvalue expressions, polynomiality and positivity\label{sec:pos}}
Our evolution formulae for the Khovanov polynomials must give the positive Laurent polynomials in $q$ and $t$, but it is far from obvious that they do. Now we rewrite the formulae so that this property becomes explicit.

In this section, we introduce the eigenvalue variables
\be
u=1,\ \ x=q^2t,\ \ y=q^4t^6.\label{evars}
\ee
as follows. First we substitute 
\be
q^2=\frac{x^4}{y},\ \ t=\frac{y}{x^3}.
\ee
Then we place proper powers of $u$ to complete the coefficients to the fractions of homogeneous polynomials, so that the total homogeneity degree is the same for all three coefficients in each series. Introduce the functions
\be
S_2(\nu;y,x)=\sgn(\nu)\frac{x^{\nu}-y^{\nu}}{x-y}=\hspace{-3mm}\sum^{\max(0,\nu)}_{\substack{i+j=\nu-1\\\min(1,\nu)}}\hspace{-3mm}x^iy^j,
\ee
and
\be
S_3(\nu;u,x,y)=\frac{u^{\nu}}{(u-x)(u-y)}+\frac{x^{\nu}}{(x-u)(x-y)}+\frac{y^{\nu}}{(y-u)(y-x)}=
\hspace{-3mm}\sum^{\max(0,\nu)}_{\substack{i+j+k=\nu-2\\\min(1,\nu)}}\hspace{-5mm}x^iy^jz^k	 
\ee
Then the reduced AAK for the positive-torus half-family can be presented in an explicitly positive-polynomial form as
\be
\Kfr^{\Tor_{n>0}}=x^{4n-2}\Big(\left(x^2+u^2+uy+y^2\right)S_2\left(\textstyle{\frac{1}{2}}(n-1);y^2,x^2\right)+S_3\left(2-n;u^2,x^2,y^2\right)\Big).
\label{Tor-eval}
\ee
A similar form for the negative-torus half-family can be obtained via the mirror symmetry
\be
\Kfr^{\Tor_{n}}(x,y,u)=\Kfr^{\Tor_{-n}}(x^{-1},y^{-1},u^{-1}).
\ee
Similarly, the AAK for the twist family can be expanded as
\be
\Kfr^{\Tw_n}=p_2x^{2\nu}+P_1S_2(\nu;u^2,x^2)+P_3S_2(\nu;y^2,x^2)+P_{13}S_3(\nu;u^2,x^2,y^2),\label{Tw-eval}
\ee
with the explicitly positive polynomial coefficients given by (\ref{Tw-EvalC}). Note that there are infinitely many expansions similar to (\ref{Tor-eval},\ref{Tw-eval}), and we just try to find the most nice one.
\be
\begin{array}{|l|c|c|}
	\hline
	n&\ge2&\le0\\
	\hline\rule{0pt}{4mm}
	\nu&\frac{2-n}{2}\le0&\frac{-n}{2}\ge0\\
	\hline\rule{0pt}{4mm}
	p_2&x^{-9}y^{-1}\left(u^2x^2+u^2y^2+x^2y^2+ux(u^2+xy)\right)&0\\
	\hline\rule{0pt}{4mm}
	P_1&u^2x^{-7}(x+y)&u^4y^{-2}(u+x)(u+y)+u^6x^{-2}+u^2x^3y^{-1}\\
	\hline\rule{0pt}{4mm}
	P_3&x^{-6}y^{-1}(u+x)(u^2+xy)&x^{-1}y^{-1}(x+u)(ux^2+yu^2+xy^2+uxy)\\
	\hline\rule{0pt}{4mm}
	P_{13}&x^{-7}(u+x)(u+y)(x+y)(u^2+xy)&u^2x^{-1}y^{-1}(u+x)^2(u+y)(x+y)\\
	\hline
\end{array}
\label{Tw-EvalC}
\ee

\section{Possible applications and further development\label{sec:conc}}
The above results broaden the class of the knot families, for which the Khovanov polynomials are explicitly described. The two-cables and the Whitehead doubles naturally complement torus, twisted, figure-eight-like, and pretzel knots. And now this list is to be further extended, probably in a more systematic way.

The used approach to the satellite polynomials is a good example of tangle calculus \cite{MMtopv,hyptangles},
which one day should become a truly working formalism for 3d TQFTs far beyond knot and Chern-Simons theories. We mean a properly extended technique which works fine for HOMFLY polynomials \cite{TanBl} 
and looks very promising for Khovanov--Rozansky \cite{KhRTang,LewLobbSl,LewLobbUp}  polynomials. This approach probably could be joined with another one, which one can call a full twist calculus \cite{NakFT,KalFT,catFT,torFT,GorFT}. Moreover, the achievements in this direction give a chance to construct a relatively simple $\Rm$-matrix-like formalism for the superpolynomials, and thus to interpret these quantities as observables in the still hypothetical {\it refined} Chern-Simons \cite{Kaul,RefCS} theory.

There is an accumulating evidence that a lot of properties of the tangle calculus 
survive for the Khovanov--Rozansky/superpolynomials. The most pronounced evidence come from the ``evolution'' formulae for knot families, which are similar for the $\Rm$-matrix polynomials and Khovanov--Rozansky/superpolynomials on each \textit{evolution domain}. The only apparent difference between the two kinds of quantities are the jumps of the latter ones on the domain \textit{walls}. However, these jumps seem to be manageable as well \cite{DMMSS,EvoDiff,AMev,DBPev,APev}. In particular, the positions of the domain walls are often governed by the long-known knot invariants, which we continue to see in the new examples above. An other common point of the supis partially liftederpolynomials and $\Rm$-matrix polynomials are differential expansions, which work quite nicely for all the known superpolynomials \cite{RasmKhR,IndSup,EvoDiff,DiffBai}. 

The above properties of the polynomials probably arise from the categorified MOY relations \cite{MOY} and semi-orthogonal decompositions for the full-twist complexes \cite{Cop,catFT,KhRrec1,KhRrec2}. Moreover, certain avatars of representation theory of quantum groups (which underlies the $\Rm$-matrix calculus) can be put in constructions for knot homologies  \cite{QueRosCol,ParabKh, AnKh}.

\section*{Acknowledgements}
A.A. is very grateful to Andrei Malyutin who patiently clarified for her all the subtleties in the definition of a satellite knot. 

\

This work was supported by the Russian Science Foundation (Grant No.21-12-00400). 

\bibliographystyle{echaya} 
\bibliography{Kh_evo}

\begin{thebibliography}{10}
\providecommand{\url}[1]{\texttt{#1}}
\providecommand{\urlprefix}{URL }
\providecommand{\selectlanguage}[1]{\relax}
\providecommand{\eprint}[2][]{\url{#2}}

\bibitem{KnotBook}
\textit{Burde G., Zieschang H., Heusener M.}
\newblock Knots.
\newblock Berlin: De Gruyter studies in mathematics, 572, 2003 P.

\bibitem{MalBelSat}
\textit{Belousov Y., Malyutin A.}
\newblock Hyperbolic knots are not generic Arxiv:\eprint{1908.06187}.

\bibitem{MalSat}
\textit{Malyutin A.}
\newblock On the question of genericity of hyperbolic knots.
\newblock International Mathematics Research Notices 2020.
\newblock V.~21 P. 7792--7828.
\newblock Arxiv:\eprint{1612.03368}.

\bibitem{KhHand}
\textit{Rose D.E., Tubbenhauer D.}
\newblock {HOMFLY-PT} homology for links in handlebodies via type {A} {Soergel}
  bimodules Arxiv:\eprint{1908.06878}.

\bibitem{AMcab}
\textit{A.Anokhina, An.Morozov}.
\newblock Cabling procedure for the colored {HOMFLY} polynomials.
\newblock Theor.Math.Phys. 2014.
\newblock V. 178 P. 1--58.
\newblock Arxiv:\eprint{1307.2216}.

\bibitem{KhoCol}
\textit{Khovanov M.}
\newblock Categorifications of the colored {Jones} polynomial.
\newblock Journal of Knot Theory and Its Ramifications 2005.
\newblock V.~14, No.~1 P. 111--130.
\newblock Arxiv:\eprint{0302060}.

\bibitem{HedCol}
\textit{Hedden M.}
\newblock Khovanov homology of the 2-cable detects the unknot.
\newblock Mathematical Research Letters 2009.
\newblock V.~16, No.~6 P. 991–994.
\newblock Arxiv:\eprint{0805.4418}.

\bibitem{MacTurCol}
\textit{Mackaay M., Turner P.}
\newblock Bar-{N}atan’s {K}hovanov homology for coloured links.
\newblock Pacific Journal of Mathematic 2007.
\newblock V. 229, No.~2 P. 429–446.
\newblock Arxiv:\eprint{0502445}.

\bibitem{BelWehCol}
\textit{Beliakova A., Wehrli S.}
\newblock Categorification of the colored {Jones} polynomial and {Rasmussen}
  invariant of links.
\newblock Canadian Journal of Mathematics 2008.
\newblock V.~60, No.~6 P. 1240 -- 1266.
\newblock Arxiv:\eprint{0510382}.

\bibitem{CapCol}
\textit{Caprau C.}
\newblock A cohomology theory for colored tangles.
\newblock Banach Center Publ. 2014.
\newblock V. 100 P. 13--25.
\newblock Arxiv:\eprint{1207.3373}.

\bibitem{ItoCol}
\textit{Ito N.}
\newblock A colored {Khovanov} bicomplex.
\newblock Banach Center Publ. 2014.
\newblock V. 103 P. 111--143.
\newblock Arxiv:\eprint{2004.08181}.

\bibitem{RozCol}
\textit{Rozansky L.}
\newblock Khovanov homology of a unicolored {B}-adequate link has a tail.
\newblock Quantum Topology 2014.
\newblock V.~4, No.~5 P. 541–579.
\newblock Arxiv:\eprint{1203.5741}.

\bibitem{WilCol}
\textit{Willis M.}
\newblock {Khovanov-Rozansky} homology for infinite multi-colored braids.
\newblock Canad. J. Math. 2020.
\newblock P. 1--39.
\newblock Arxiv:\eprint{1904.09055}.

\bibitem{RosWedCol}
\textit{Rose D.E.V., Wedrich P.}
\newblock Deformations of colored {sl(N)} link homologies via foams.
\newblock Geometry \& Topology 2016.
\newblock V.~20 P. 3431–3517.
\newblock Arxiv:\eprint{1501.02567}.

\bibitem{QueRosCol}
\textit{Queffelec H., Rose D.E.V.}
\newblock The {$\mathfrak{sl}_n$} foam 2-category: a combinatorial formulation
  of {K}hovanov-{R}ozansky homology via categorical skew {H}owe duality.
\newblock Advances in Mathematics 2016.
\newblock V. 302 P. 1251--1339.
\newblock Arxiv:\eprint{1405.5920}.

\bibitem{RobWagcol}
\textit{Robert L.H., Wagner E.}
\newblock A closed formula for the evaluation of {$\mathfrak{sl}_n$}-foams.
\newblock Quantum Topology 2020.
\newblock V.~11, No.~3 P. 411–487.
\newblock Arxiv:\eprint{1702.04140}.

\bibitem{LukHomCal}
\textit{Lewark L.}
\newblock Lukas lewark homepage. {Knot} software.
\newblock \urlprefix\url{http://www.lewark.de/lukas/software.html}.

\bibitem{RobWag}
\textit{Robert L.H., Wagner E.}
\newblock {A} quantum categorification of the {A}lexander polynomial 2019.
\newblock Arxiv:\eprint{1902.05648}.

\bibitem{SlepNovAl}
\textit{Mishnyakov V., Sleptsov A., Tselousov N.}
\newblock A novel symmetry of colored {HOMFLY} polynomials coming from
  {$\mathfrak{sl}$(N|M)} superalgebras 2020.
\newblock Arxiv:\eprint{2005.01188}.

\bibitem{HedWh}
\textit{Hedden M.}
\newblock Knot {F}loer homology of {W}hitehead doubles.
\newblock Geom. Topol. 2007.
\newblock V.~11 P. 2277--2338.
\newblock Arxiv:\eprint{0606094}.

\bibitem{HedCab1}
\textit{Hedden M.}
\newblock On knot {F}loer homology and cabling.
\newblock Geom. Topol. 2005.
\newblock V.~5 P. 1197--1222.
\newblock Arxiv:\eprint{0406402}.

\bibitem{HedCab2}
\textit{Hedden M.}
\newblock On knot {F}loer homology and cabling {II}.
\newblock IMRN 2008.
\newblock V. 2009, No.~12 P. 2248–2274.
\newblock Arxiv:\eprint{0806.2172}.

\bibitem{MorTwSat}
\textit{Morozov A.}
\newblock Knot polynomials for twist satellites.
\newblock Phys.Lett. 2018.
\newblock V. B782 P. 104--111.
\newblock Arxiv:\eprint{1801.02407}.

\bibitem{HOMFLY}
\textit{Freyd P., Yetter D., Hoste J., Lickorish W.B.R., Millett K., Ocneanu
  A.}
\newblock A new polynomial invariant of knots and links.
\newblock Bull. AMS 1985.
\newblock V.~12 P. 239--246.

\bibitem{PT}
\textit{Przytycki J.H., Traczyk P.}
\newblock Invariants of links of {Conway} type.
\newblock Kobe J. Math. 1988.
\newblock V.~4 P. 115--139.

\bibitem{Kaul}
\textit{Kaul R.K.}
\newblock {Chern-Simons} theory, knot invariants, vertex models and
  three-manifold invariants.
\newblock Frontiers of field theory, quantum gravity and strings. Proceedings
  1999.
\newblock P. 45--63.
\newblock Arxiv:\eprint{9804122}.

\bibitem{ReshTur}
\textit{Reshetikhin N.Y., Turaev V.G.}
\newblock Ribbon graphs and their invariants derived from quantum groups.
\newblock Commun. Math. Phys. 1990.
\newblock V. 127 P. 1--26.

\bibitem{MorSm}
\textit{Morozov A., Smirnov A.}
\newblock {Chern-Simons} theory in the temporal gauge and knot invariants
  through the universal quantum {R-matrix}.
\newblock Nucl. Phys. 2010.
\newblock V. B835 P. 284--313.
\newblock Arxiv:\eprint{1001.2003}.

\bibitem{TabArb}
\textit{Mironov A., Morozov A., Morozov A., Ramadevi P., Singh V.K., Sleptsov
  A.}
\newblock Tabulating knot polynomials for arborescent knots.
\newblock J. Phys. A: Math. Theor. 2017.
\newblock V.~50, No. 085201.
\newblock Arxiv:\eprint{1601.04199}.

\bibitem{MirMorUn}
\textit{A.Mironov, A.Morozov}.
\newblock Universal {Racah} matrices and adjoint knot polynomials. {I}.
  {Arborescent} knots.
\newblock Physics Letters 2016.
\newblock V. B755 P. 47--57.
\newblock Arxiv:\eprint{1511.09077}.

\bibitem{DunGukRas}
\textit{Dunfield N.M., Gukov S., Rasmussen J.}
\newblock The superpolynomial for knot homologies.
\newblock Experimental Math. 2006.
\newblock V.~15 P. 129--159.
\newblock Arxiv:\eprint{0505662}.

\bibitem{KhR}
\textit{Khovanov M., Rozansky L.}
\newblock Matrix factorizations and link homology.
\newblock Fund. Math. 2008.
\newblock V. 199 P. 1--91.
\newblock Arxiv:\eprint{0401268}.

\bibitem{Gor4gr}
\textit{Gorsky E., Gukov S., Stosic M.}
\newblock Quadruply-graded colored homology of knots 2014.
\newblock Arxiv:\eprint{1304.3481}.

\bibitem{Arth4gr}
\textit{S.Arthamonov, A.Mironov, A.Morozov}.
\newblock Differential hierarchy and additional grading of knot polynomials.
\newblock Theor.Math.Phys. 2014.
\newblock V. 179 P. 509--542.
\newblock Arxiv:\eprint{1306.5682}.

\bibitem{katlas}
\textit{Bar-Natan D., Scott M., et~al.}
\newblock The {K}not {A}tlas.
\newblock \urlprefix\url{http://katlas.org}.

\bibitem{HedOrd}
\textit{Hedden M., Ording P.}
\newblock The {Ozsv\'ath-Szab\'o} and rasmussen concordance invariants are not
  equal.
\newblock American Journal of Mathematics 2008.
\newblock V. 13011, No.~2 P.~2.
\newblock Arxiv:\eprint{0512348}.

\bibitem{knbook}
\textit{Morozov A., Sleptsov A., et~al.}
\newblock The knotebook.
\newblock \urlprefix\url{www.knotebook.org}.

\bibitem{DBPev}
\textit{Dunin-Barkowski P., Popolitov A., Popolitova S.}
\newblock Evolution for {Khovanov} polynomials for figure-eight-like family of
  knots Arxiv:\eprint{1812.00858}.

\bibitem{APev}
\textit{Anokhina A., A.Morozov, A.Popolitov}.
\newblock Nimble evolution for pretzel {Khovanov} polynomials.
\newblock Eur. Phys. J. C 2019.
\newblock V.~79, No. 867.
\newblock Arxiv:\eprint{1904.10277}.

\bibitem{MMtopv}
\textit{Awata H., Kanno H., Mironov A., Morozov A., Morozov A.}
\newblock A non-torus link from topological vertex.
\newblock Phys. Rev. 2018.
\newblock V. D 98, No. 046018.
\newblock Arxiv:\eprint{1806.01146}.

\bibitem{hyptangles}
\textit{Awata H., Kanno H., Mironov A., Morozov A.}
\newblock Can tangle calculus be applicable to hyperpolynomials?
\newblock Nuclear Physics B 2019.
\newblock V. 949, No. 114816.
\newblock Arxiv:\eprint{1905.00208}.

\bibitem{TanBl}
\textit{Mironov A., Morozov A., Morozov A.}
\newblock Tangle blocks in the theory of link invariants.
\newblock JHEP 2018.
\newblock V. 2018 P. 128.
\newblock Arxiv:\eprint{1804.07278}.

\bibitem{KhRTang}
\textit{Krasner D.}
\newblock A computation in {Khovanov-Rozansky} homology.
\newblock Fund.Math. 2009.
\newblock V. 203, No.~1 P. 75--95.
\newblock Arxiv:\eprint{0801.4018}.

\bibitem{LewLobbSl}
\textit{Lewark L., Lobb A.}
\newblock New quantum obstructions to sliceness.
\newblock Proceedings of the London Mathematical Society 2016.
\newblock V. 112, No.~1 P. 81--114.
\newblock Arxiv:\eprint{1501.07138}.

\bibitem{LewLobbUp}
\textit{Lewark L., Lobb A.}
\newblock Upsilon-like concordance invariants from sl(n) knot cohomology.
\newblock Geom. Topol. 2019.
\newblock V.~23 P. 745--780.
\newblock Arxiv:\eprint{1707.00891}.

\bibitem{NakFT}
\textit{Nakagane K.}
\newblock A full-twisting formula for the {HOMFLY} polynomial
  Arxiv:\eprint{2009.05511}.

\bibitem{KalFT}
\textit{K\`alm\`an T.}
\newblock Meridian twisting of closed braids and the {HOMFLY} polynomial.
\newblock Proc. Cambridge Philos. Soc 2009.
\newblock V. 146, No.~3 P. 649–660.
\newblock Arxiv:\eprint{0803.0103}.

\bibitem{catFT}
\textit{Elias B., Hogancamp M.}
\newblock Categorical diagonalization of full twists Arxiv:\eprint{1801.00191}.

\bibitem{torFT}
\textit{Nakagane K.}
\newblock The action of full twist on the superpolynomial for torus knots.
\newblock Topology and its Applications 2019.
\newblock V. 266, No. 106841.
\newblock Arxiv:\eprint{1805.01606}.

\bibitem{GorFT}
\textit{Gorsky E., Hogancamp M., Mellit A., Nakagane K.}
\newblock Serre duality for {Khovanov-Rozansky} homology.
\newblock Selecta Mathematica 2019.
\newblock V.~25, No.~79.
\newblock Arxiv:\eprint{1902.08281}.

\bibitem{RefCS}
\textit{Aganagic M., Shakirov S.}
\newblock Knot homology from refined {Chern-Simons} theory
  Arxiv:\eprint{1105.5117}.

\bibitem{DMMSS}
\textit{Dunin-Barkowski P., Mironov A., Morozov A., Sleptsov A., Smirnov A.}
\newblock Superpolynomials for torus knots from evolution induced by
  cut-and-join operators.
\newblock JHEP 2013.
\newblock V.~03, No. 021.
\newblock Arxiv:\eprint{1106.4305}.

\bibitem{EvoDiff}
\textit{Mironov A., Morozov A., Morozov A.}
\newblock Evolution method and ``differential hierarchy'' of colored knot
  polynomials.
\newblock AIP Conf. Proc. 2013.
\newblock V. 1562.
\newblock Arxiv:\eprint{1306.3197}.

\bibitem{AMev}
\textit{A.Anokhina, A.Morozov}.
\newblock Are {Khovanov-Rozansky} polynomials consistent with evolution in the
  space of knots?
\newblock JHEP 2018.
\newblock V. 1804, No. 066.
\newblock Arxiv:\eprint{1802.09383}.

\bibitem{RasmKhR}
\textit{Rasmussen J.}
\newblock Some differentials on {Khovanov-Rozansky} homology 2006.
\newblock Arxiv:\eprint{0607544}.

\bibitem{IndSup}
\textit{Nawata S., Ramadevi P., Zodinmawia}.
\newblock Colored {Kauffman} homology and super-{A}-polynomials.
\newblock JHEP 2014.
\newblock V. 1401, No. 126.
\newblock Arxiv:\eprint{1310.2240}.

\bibitem{DiffBai}
\textit{Bai C., Jiang J., Liang J., Mironov A., Morozov A., Morozov A.,
  Sleptsov A.}
\newblock Differential expansion for link polynomials.
\newblock Phys.Lett. 2018.
\newblock V. B778 P. 197--206.
\newblock Arxiv:\eprint{1709.09228}.

\bibitem{MOY}
\textit{Murakami H., Ohtsuki T., Yamada S.}
\newblock {HOMFLY} polynomial via an invariant of colored plane graphs.
\newblock Enseign. Math. 1998.
\newblock V. 2(44), No. 3--4 P. 325–360.

\bibitem{Cop}
\textit{Cooper B., Krushkal V.}
\newblock Categorification of the {Jones-Wenzl} projectors.
\newblock Quantum Topol. 2012.
\newblock V.~3 P. 139--180.
\newblock Arxiv:\eprint{1005.5117}.

\bibitem{KhRrec1}
\textit{Elias B., Hogancamp M.}
\newblock On the computation of torus link homology.
\newblock Compositio Mathematica 2019.
\newblock V. 155, No.~1 P. 164--205.
\newblock Arxiv:\eprint{1603.00407}.

\bibitem{KhRrec2}
\textit{Hogancamp M.}
\newblock {Khovanov-Rozansky} homology and higher {Catalan} sequences 2017.
\newblock Arxiv:\eprint{1704.01562}.

\bibitem{ParabKh}
\textit{Elias B., Khovanov M.}
\newblock Diagrammatics for {Soergel} categories.
\newblock Journal of Mathematics and Mathematical Sciences 2010.
\newblock V. 2010, No. 978635.
\newblock Arxiv:\eprint{0902.4700}.

\bibitem{AnKh}
\textit{Anokhina A.}
\newblock Towards formalization of the soliton counting technique for the
  {Khovanov-Rozansky} invariants in the deformed {R}-matrix approach.
\newblock ATMP 2018.
\newblock V. 33:6, No. 1850221.
\newblock Arxiv:\eprint{1710.07306}.

\bibitem{KlimSch}
\textit{Klimyk A., Schm{\"u}dgen K.}
\newblock Quantum groups and their representations.
\newblock Berlin Heidelberg: Springer, 2012, 552 P.

\bibitem{DM3}
\textit{Dolotin V., Morozov A.}
\newblock Introduction to {Khovanov} homologies. {III.} {A} new and simple
  tensor-algebra construction of {Khovanov-Rozansky} invariants.
\newblock Nucl. Phys. 2014.
\newblock V. B878 P. 12--81.
\newblock Arxiv:\eprint{1308.5759}.

\bibitem{indknot}
\textit{Cha J.C., Livingston C.}
\newblock {KnotInfo}: Table of knot invariants.
\newblock \urlprefix\url{http://www.indiana.edu/~knotinfo}.

\end{thebibliography}

\appendix
\section{Cabling and an invariant definition of the $m$-strand satellite\label{app:FT}}
In this section, we assume that a satellite is defined by a planar diagram, like one in fig.\ref{fig:sat}. We wish to substitute the two strands that first go along the companion and than intertwine with each other with any number of such ones.

Let $\Km_w^{\circ m}$ be the satellite of the knot $\Km$ that is obtained when each line on the $\Km$ diagram with the writhe number $w$ is substituted with $m$ parallel strands. The satellite polynomial is then expanded over the coloured polynomials of the as 
\be
\check{H}^{\Km_{\nu}^{\circ m}}=\sum_{Q\vdash \square^{\otimes m}}\check{H}_Q^{\Km_{\nu}},\label{m-cab-vert}
\ee
where the colour (representation) label $Q$ runs over all partitions of the $m$. Identity (\ref{m-cab-vert}) holds for the unreduced polynomials in the group theory normalisation (in the \textit{vertical framing})~\cite{AMcab}, which are defined for a knot diagram and are knot invariants only up to a factor. The true unreduced knot polynomials (in the \textit{topological framing}) differ by the exponential of the writhe number, namely 
\be H^{\Km}_Q=A^{-m\nu}q^{-2\varkappa_Q\nu}\check{H}_Q^{\Km_\nu},\ee
where $Q$ is a partition of $m$ and $\varkappa_Q$ is an explicitly defined integer-valued function of $Q$ (the second Casimir of the representation $Q$~\cite{KlimSch}). In particular, the plain (uncoloured) HOMFLY is associated with the partition $Q=1$ of $m=1$ and has $\varkappa=0$. Hence, the analogue of (\ref{m-cab-vert}) for the topological invariant quantities is 
\be 
A^{m^2\nu}H^{\Km_{\nu}^{\circ m}}=\sum_{Q\vdash \square^{\otimes m}}A^{m\nu}q^{2\varkappa_Q\nu}\Hm_Q^{\Km_{\nu}}\ \ \Rightarrow\ \
\Hm^{\Km_{\nu}^{\circ m}}=\sum_{Q\vdash \square^{\otimes m}}\Big(A^{m(m-1)}q^{2\varkappa_Q}\Big)^{\nu}\Hm_Q^{\Km}\label{m-cab-top}
\ee
where we substituted the writhe number $W=m^2w$ of the $\Km_w^{\circ m}$ defined above (see examples in fig.\ref{fig:sat} where $m=2$). We omit the label $w$ in $H_Q^{\Km}$, since they depend only on the knot $\Km$, unlike the $\Km_w^{\circ m}$, whose definition still depends on the knot diagram.

On the other hand, the full twist on $m$ strands, which is represented by the braid word  
\be FT_m=(m\!-\!1)\ldots21(m\!-\!2)\ldots21\ldots\ldots1,\ee is a distinguished braid group element~\cite{torFT,KalFT}. I.e., insertion of the $\nu$ copies of the full twists in the $m$-strand cable results just in an extra factor of the form $c_Q^{\nu}$. Precisely, $\check{c}_Q^{\nu}=q^{-2\varkappa_Q}$ in the vertical framing, as follows from the Rosso--Jones formula (e.g., (11) of \cite{AMev}), so that $\check{c}_Q^{\nu}= A^{-m(m-1)}q^{-2\varkappa_Q}$ in the topological framing (see (15) of \cite{AMev}), because the $FT_m$ contains $m(m-1)$ co-oriented crossings. Hence, (\ref{m-cab-top}) can be rewritten as
\be 
\Hm^{\Km_{\nu}^{\circ m}}=\sum_{Q\vdash \square^{\otimes m}}\Hm_Q^{FT^{-\nu}_m\,\sharp\,\Km_{\nu}}
\ee
\be
\Hm^{\Sm_{\Tor^m_{\kappa}}^{\Km}}=\sum_{Q\vdash \square^{\otimes m}}\mu^{\kappa}\Hm_Q^{\Km},\ \ \mbox{for}\ 
\Sm_{\Tor^m_{\kappa}}^{\Km}=FT^{\kappa-{\nu}}_m\,\sharp\,\Km_{\nu},\label{FT-m-cab-top}
\ee
where
\be \mu=A^{-m(m-1)}q^{-2\varkappa_Q}.\ee
and the notation $FT^{w^{\prime}}_m\sharp\Km_w$ implies that we insert $w^{\prime}>0$ or $(-\!w^{\prime})>0$ copies of the $FT_m$ or of its mirror image $\overline{\rule{0pt}{3.5mm}FT}_m$, respectively, in \textit{a} section of the $m$ strand cable between any two groups of crossings that substitute a  crossing in $\Km$.  

The coloured polynomials $H_Q^{\Km}$ in the r.h.s. of (\ref{FT-m-cab-top}) are the knot invariants, and the coefficients $\mu$ do not depend on the knot at all. Hence, the above defined $\Sm_{\Tor^m_{\kappa}}^{\Km}$ is a topological invariant as long as so does the $k$.

If $m=2$, then the satellite is the torus satellite $\Sm_{\Tor^2_{\kappa}}^{\Km}\equiv\Sm_{\Tor_k}^{\Km}$ we study above. The colour label in (\ref{m-cab-vert}) then runs over the partitions of $m=2$, which are $Q=[2]$ and $Q=[1,1]$ and have $\varkappa_{[2]}=2$ and $\varkappa_{[1,1]}=-1$. Hence (\ref{m-cab-vert}) and (\ref{m-cab-top}) are reduced to (\ref{tor-cab-vert-u}) and (\ref{tor-cab-top-u}), respectively. The pattern includes $w$ half-twists on the two strands. Hence, one can set $\kappa-\nu=w/2$ for the number of the full

\begin{table}
	\caption{Basic properties of the knot polynomials we study\label{tab:pols}.}
	$$
	\arraycolsep=0.5mm
	\begin{array}{ccccccccccccccc}
	&&&&\overline{\rule{0pt}{3mm}\Tor}_n&=&\Tor_{-n}\\
	&&&&\overline{\rule{0pt}{3mm}\Tw}^{\circ\circ}_n&=&\Tw^{\bullet\bullet}_{-n}\\
	&&&&\overline{\rule{0pt}{3mm}\Tw}^{\bullet\bullet}_n&=&\Tw^{\circ\circ}_{-n}
	\\[2mm]
	\sigma^{\Km}(A)&=&\sigma^{\bar{\Km}}(\frac{1}{A})&\stackrel{q\!=\!1}{\longleftarrow}&\Hm^{\Km}(A,q)&=&\Hm^{\bar{\Km}}(\frac{1}{A},q^{-1})&\stackrel{A\!=\!q^2}{\longrightarrow}
	&J^{\Km}(q)&=&J^{\bar{\Km}}(\frac{1}{q})&\stackrel{t\!=\!-1}{\longleftarrow}&\Kh^{\Km}(q,t)&=&\Kh^{\bar{\Km}}(\frac{1}{q},\frac{1}{t})
	\\[-2mm]
	&&&&&\circlearrowleft\\[-2mm]&&&&\boxed{q\leftrightarrow-\textstyle{\frac{1}{q}}}\\[-1mm]&&&&&\circlearrowleft\\[-2mm]
	&&&&\Hm_{\adj}^{\Km}(A,q)&=&\Hm_{\adj}^{\bar{\Km}}(\frac{1}{A},\frac{1}{q})\\[-4mm]
	&&&\swarrow&&&&\searrow\\
	\multicolumn{3}{c}{\big[\sigma^{\Km}(A)\big]^2}&\stackrel{q\!=\!1}{\longleftarrow}&
	\Hm_{\sym}^{\Km}(A,q)&=&\Hm_{\sym}^{\bar{\Km}}(\frac{1}{A},\frac{1}{q})
	&\stackrel{A\!=\!q^2}{\longrightarrow}
	&J_{\begin{array}{c}\\[-3mm]\sym\\[-1mm](\adj)\end{array}}^{\Km}\hspace{-7mm}(q)&=&
	J_{\begin{array}{c}\\[-3mm]\sym\\[-1mm](\adj)\end{array}}^{\bar{\Km}}\hspace{-7mm}(\frac{1}{q})&\stackrel{t\!=\!-1}{\longleftarrow}&
	\multicolumn{3}{c}{"\ \Kh^{\Km}_{\begin{array}{c}\\[-3mm]\sym\\[-1mm](\adj)\end{array}}\hspace{-7mm}(q,t)\ "}\\[-6mm]
	&&&&\boxed{q\leftrightarrow-\textstyle{\frac{1}{q}}}&\updownarrow\\[2mm]
	&&&&\Hm_{\asm}^{\Km}(A,q)&=&\Hm_{\asm}^{\bar{\Km}}(\frac{1}{A},\frac{1}{q})&\\[-3mm]
	&&&&&&&\hspace{-5mm}\stackrel{\hspace{5mm}A=q^2}{\searrow}&\\
	&&1&=&\multicolumn{3}{c}{\Hm^{\Km}_{\trv}(A,q)}&=&\multicolumn{3}{c}{J^{\Km}_{\trv}(q)}&=&\multicolumn{3}{c}{\Kh^{\Km}_{\trv}(q,t)}
	\end{array}
	$$
\end{table}

\section{First symmetric polynomials for the simplest knot families}
Here we summarise the explicit formulae and the basic properties of the coloured HOMFLY polynomials of the two-strand torus and twist knots. The need references can be found in \cite{knbook}.

\subsection{Two-strand torus knots\label{app:coltor}}
All torus knots $Tor_n$ (fig.\ref{fig:tortw}.I) have odd half-twist number $n$ (while even $n$ yields a two-component link). The first torus knots have the standard prime-knot names $n_1$ ($n=3,5,7,9$), and $11_{a367}$ ($n=-11$).
The torus knots with $n$ and $-n$ are  the mirror images of each other, in general the topologically distinct knots. The exceptions are $\Tor_{-1}=\Tor_1$, which both represent the unknot. Hence the Jones, HOMFLY, and Khovanov polynomials equal 1 for $\Tw_{\pm 1}$.
\be\arraycolsep=1mm
\begin{array}{|c|c|c|}
	\hline
	\multicolumn{3}{|c|}{\rule{0pt}{6mm}\text{Torus knots},\ \Tor_n,\ n=2k-1}\\
	\multicolumn{3}{|c|}{\rule{0pt}{6mm}\ \Hm^{\Tor_n}=\sum_{\lambda}C_{\lambda}\lambda^n}\\
	\hline\rule{0pt}{7mm}
	\begin{tabular}{c}Eigen-\\values,\end{tabular}&\multicolumn{2}{c|}{\text{Coefficients},C_{\lambda}}\\\cline{2-3}
	\lambda&\text{Symmetric representation}&\text{Adjoint representation}\rule{0pt}{6mm}\\[2mm]
	\hline\rule{0pt}{6mm}
	(A/q)^{-2}
	&\frac{\left\{Aq^2\right\}\left\{Aq^3\right\}}{\left\{q^3\right\}\left\{q^4\right\}}
	&\frac{\left\{A\right\}^2\left\{Aq^3\right\}}{\left\{q^2\right\}\left\{Aq\right\}}\\[2mm]
	\hline\rule{0pt}{6mm}
	-(Aq)^{-2}
	&\frac{\left\{Aq^2\right\}\left\{Aq^{-1}\right\}}{\left\{q\right\}\left\{q^4\right\}}
	&\frac{\left\{A\right\}^2\left\{Aq^{-3}\right\}}{\left\{q^2 \right\}\left\{Aq^{-1}\right\}}\\[2mm]
	\hline\rule{0pt}{6mm}
	-(Aq^2)^{-2}
	&\frac{\left\{A\right\}\left\{Aq^{-1}\right\}}{\left\{q^2\right\}\left\{q^3\right\}}&-\\[2mm]
	\hline\rule{0pt}{6mm}
	-A^{-2}&-&\begin{array}{c}\\[2mm]\frac{\left\{Aq^2\right\}\left\{Aq^{-2}\right\}}{\left\{q^2\right\}^2}\end{array}\\[2mm]
	\cline{1-2}\rule{0pt}{6mm}
	-A^{-2}&-&\\[2mm]	
	\hline
	\rule{0pt}{6mm}
	A^{-4}&-&\frac{\left\{q\right\}^2}{\{Aq\}\{Aq^{-1}\}}\\[2mm]	
	\hline
\end{array}
\label{colTor}
\ee

\subsection{Twist knots\label{app:coltw}}
The twist knots are obtained both for odd and even half-twists numbers $n$, but the knot diagrams with $2k-1$ and $2k$ half twists yield the same knot. Yet there is no nice formula for the HOMFLY polynomials where $n$ can be both odd and even, but rather two separate formulae for the two cases. Unlike the case of torus knots, the mirror image of a twist knot $\Tw_n$ the topologically distinct from the knot $\Tw_{-n}$. Instead that, there are the two different families of the knots $\Tw^{\circ\circ}$ and $\Tw^{\bullet\bullet}$, differing by the orientation of the two lock-down crossings (adjacent to the gray area in fig.\ref{fig:tortw}.II). The mirror symmetry maps one family into the other~\cite{DM3}. The two exceptions are the unknot $0_1=\bar{\Tw^{\circ\circ}}_0=\Tw^{\bullet\bullet}_0$ and the figure eight knot~\cite{katlas} $4_1=\bar{\Tw^{\circ\circ}}_{-2}=\Tw^{\bullet\bullet}_{2}$. Each of them belongs to the both families and is the mirror image of itself. This differs the twist knots from the two-strand torus knots, where the knot $\Tor_n$ and its mirror image $\overline{\rule{0pt}{3mm}\Tor}_n=\Tor_{-n}$ belong to the same family. Here we write $\Tw_n$ assuming that the lock-down crossing are co-oriented with other ones for $n>0$ (fig.\ref{fig:tortw}.II). Hence $n=2,4,6,8$ and $10$ yields mirrors of the knots $3_1$,$5_2$,$7_2$, $9_2$, and the knot $11_{a247}$, while $n=-2,-4,-6,-8$ yields the knots $4_1,6_1,8_1,10_1$ from \cite{katlas}.

The reduced Jones, HOMFLY and Khovanov polynomials equal 1 for the unknot $\Tw_0$.

\be
\arraycolsep=0mm
\begin{array}{|c|c|c|}
	\hline
	\multicolumn{3}{|c|}{\rule{0pt}{6mm}\text{Twist knots},\ \Tw_n,\ n=2k}\\
	\multicolumn{3}{|c|}{\rule{0pt}{6mm}\ \Hm^{\Tw_n}=\sum_{\lambda}C_{\lambda}\lambda^n}\\
	\hline\rule{0pt}{7mm}
	\begin{tabular}{c}Eigen-\\values,\end{tabular}&\multicolumn{2}{c|}{\text{Coefficients},C_{\lambda}}\\\cline{2-3}
	\lambda&\text{Symmetric representation}&\text{Adjoint representation}\rule{0pt}{6mm}\\[2mm]
	\hline\rule{0pt}{6mm}
	1&\ 1\!+\!\frac{\left\{Aq^{-1}\right\}\left\{Aq^2\right\}\left(1-(1-q^2)(1-q^4)A^2-A^4q^4\right)}{A^4q^4\left\{A\right\}\left\{Aq\right\}}&\
	1\!-\!\frac{A\left\{A\right\}}{\left\{Aq\right\}\left\{Aq^{-1}\right\}}\!\!\left(\!\!
	\left(\!\frac{q^3+q^{-3}}{q+q^{-1}}\!\right)^2\!\!\!A^{-6}\!-\!\frac{q^5+q^{-5}}{q+q^{-1}}A^{-4}\!-\!\frac{q^3+q^{-3}}{q+q^{-1}}A^{-2}\!+\!1\!
	\right)\\[2mm]
	\hline\rule{0pt}{6mm}
	A^{-1}&-\frac{(1+q^2)\left(1-A^2+A^2q^2-A^2q^6\right)\left\{Aq^{-1}\right\}}{A^4q^5\left\{A\right\}}
	&\begin{array}{c}\\[2mm]
		\frac{\left\{A\right\}\left(q^3+q^{-3}\right)}{A^3\left(q+q^{-1}\right)}
	\end{array}\\[2mm]
	\cline{1-2}\rule{0pt}{6mm}
	-A^{-1}&-&\\[2mm]
	\hline\rule{0pt}{6mm}
	(Aq)^{-2}&\frac{\left\{A\right\}\left\{Aq^3\right\}\left\{Aq^{-1}\right\}}{qA^2\left\{Aq\right\}}
	&
	\frac{\left\{A\right\}^2\left\{Aq^3\right\}\left\{A^2q^{-2}\right\}}{\left(q+q^{-1}\right)^2\left\{Aq\right\}\left\{Aq^{-1}\right\}} \\[2mm]
	\hline\rule{0pt}{6mm}
	(A/q)^{-2}
	&-&\frac{\left\{A\right\}^2\left\{Aq^{-3}\right\}\left\{A^2q^2\right\}}{A^3q\left(q+q^{-1}\right)^2\left\{Aq\right\}\left\{Aq^{-1}\right\}}\\[2mm]
	\hline\rule{0pt}{6mm}
	-A^{-2}&-&\begin{array}{c}\\[2mm]\frac{\left\{A\right\}\left\{Aq^2\right\}\left\{Aq^{-2}\right\}}{A^3\left(q+q^{-1}\right)^2}\end{array}\\[2mm]
	\cline{1-2}\rule{0pt}{6mm}
	-A^{-2}&-&\\[2mm]	
	\hline
\end{array}
\label{colTw}\ee

\section{Technical guidance to the experimental data\label{app:data}}

\subsection{Simplest knots}
The source files have the names like ``kh-red-precomp-whiteheadized-rolfsen-knot-c-m'', where ``c-m'' is the Rolfsen name of the knot \cite{katlas}, e.g., ``6-3''. The MAPLE function is called ``KhTwKn(kn,k)'', where the $kn$ is the knot number in the list ``Knots'' (contains all available cases) and $k$ is the satellite class (see sec.\ref{sec:sat}). 
\be
\arraycolsep=0.5mm
\begin{array}{|c|c|c|c|c|c|c|c|c|c|c|c|c|c|c|c|c|c|c|c|c|c|c|c|}
\multicolumn{24}{c}{(q^2t)^{-1}\cdot \mathrm{PrecompKhRed}[\Km,\Cr_{\Km}+k,2]=\Kh^{\Sm_{\Tw_k}^{\Km}}}\\[1mm]
\hline\rule{0pt}{4mm}
\Km&3_1&4_1&5_1&5_2&6_1&6_2&6_3&7_1&7_2&7_3&7_4&7_5&7_6&7_7&8_1
&8_2&8_3&8_4&8_5&8_6&8_7&8_8&8_9\\
\hline
\simeq&\Tor_{3}&&\Tor_5&&&&&\Tor_7&&&&&&&&&&&&&&&\\
\hline
\simeq&\rule{0pt}{3mm}\overline{\rule{0pt}{3mm}\Tw}_{2}&\Tw_{-2}&&\Tw_3&\Tw_{-4}&&&&\Tw_5&&&&&&\Tw_{-6}&&&&&&&&\\
\hline
\Cr_{\Km}&0&0&-2&4&4&-2&0&-4&8&-2&-8&2&0&-2&8&-4&0&-6&4&2&2&-4&0\\
\hline\hline
\multicolumn{6}{|c|}{\rule{0pt}{4mm}\Cr_{\Tor_{n<0}}=3n+3}&\multicolumn{6}{c|}{\Cr_{\Tor_{n>0}}=3n-3}&
 \multicolumn{6}{c|}{\Cr_{\Tw_{n\le0}}=0}&\multicolumn{6}{c|}{\Cr_{\Tw_{n>0}}=-6}\\[0.5mm]
\hline
\end{array}
\label{RolfData}
\ee
Note that the Rolfsen notation is related to one knot from each pair of mirror images, and different images can be chosen in different knot tables. In particular, the choice in \cite{katlas,knbook} and \cite{indknot} is the same for knots $7_3$,$7_4$,$8_4$,$8_5$ and opposite for all other knots up to $8_9$.
\subsection{Simplest knot families}
The source files have the names like ``kh-red-precomp-$<$label$>$-$<$knot name$>$'' (reduced polynomials) or ``kh-precomp-$<$label$>$-$<$knot name$>$'' (unreduced polynomials, the torus-torus case). The notations are explained in sec.\ref{sec:knoque}. 
\be
\arraycolsep=1mm
\begin{array}{|p{2cm}|p{2cm}||p{3.8cm}|p{3.5cm}|p{1.8cm}|}
	\hline\multicolumn{2}{|p{6cm}||}{\rule{0pt}{6mm}Description of families}&\multicolumn{2}{p{6cm}|}{\rule{0pt}{6mm}Matching pairs}\\	
	\hline\rule{0pt}{6mm}Companion,&$\begin{array}{p{1.5cm}}Pattern/\\companion,\end{array}$MAPLE&Source file, label;&&source/\\
	$\sharp$half-twists,&$\sharp$half-twists&function&the matching&MAPLE\\
	writhe&&
	(arguments)&arguments&factor\\
	\hline\rule{0pt}{6mm}	
	Torus&Torus,&KuTorTor \mbox{$(-\!n,\!-\!n\!+\!3\sgn_n\!+\!w)$}&twst-torus&\multicolumn{1}{c|}{q^2t,n\!>\!0, red.}\\
	$n\!=\!2l\!+\!1$,&$w\!=\!\!2p\!+\!1$&KTorTor \mbox{$(\!n,\!\!n\!-\!3\sgn_n\!+\!w)$} 
	&\multicolumn{1}{c|}{n,w}&
	\multicolumn{1}{c|}{q^{-1}, \text{ othw.}}\\
	\cline{2-2}\cline{3-5}\rule{0pt}{5mm}$\nu=n$&Twist, &KTorTw \mbox{$(\!n,\!\!n\!-\!3\sgn_n+\!w)$}&whiteheadized-torus&\\
	&$w\!=\!\!2p\!$
	& 
	&\multicolumn{1}{c|}{n\!<\!0,w;\ n\!>\!0,w,2}&\multicolumn{1}{c|}{q^2t}\\
	\hline\rule{0pt}{6mm}
	Twist,&Torus,&KTwTor \mbox{$(-n,2n\!+\!w\!-\!4\!+\!6\Theta_{-n})$}&twisted-two-strand&\\
	$n\!=\!2l$&$w\!=\!\!2p\!+\!1$
	& 
	&\multicolumn{1}{c|}{n,w}&\multicolumn{1}{c|}{q^{-1}}\\
	\cline{2-2}\cline{3-5}\rule{0pt}{5mm}$\nu=n+2$&Twist, &KTwTw \mbox{$(-\!n,2n\!+\!w\!-\!4\!+\!6\Theta_{-n})$}&twisted-twisted&\\
	&$w\!=\!\!2p\!$& 
	&\multicolumn{1}{c|}{n,w}&\multicolumn{1}{c|}{q^2t}\\
	\hline
\end{array}
\label{TorTwData}
\ee

\subsection{Precaution}

The program from \cite{katlas} seems to work incorrectly for some knot diagrams. The simplest example is the diagram $X[1,2,2,1]$ that represents the twisted unknot. This can be usually coped with by inserting a long enough trivial two-strand  braid of the form $(1,-1,1,-1,\ldots)$ (converted to a Gauss diagram).


\end{document}